\documentclass[aps,prd,preprint,groupedaddress,showpacs]{revtex4}
\usepackage{epsfig}
\usepackage{graphics}
\usepackage{slashed}
\usepackage{color}
\begin{document}

\leftmargin -2cm
\def\choosen{\atopwithdelims..}




\title{Prompt $\Upsilon(nS)$ production at the LHC in the Regge
limit of QCD}

\author{\firstname{M.A.
}\surname{Nefedov}}\email{nefedovma@gmail.com}
\author{\firstname{V.A.} \surname{Saleev}}
\email{saleev@samsu.ru}
\author{\firstname{A.V. }\surname{Shipilova}}\email{alexshipilova@samsu.ru}

 \affiliation{ Samara
State University, Academic Pavlov Street 1, 443011 Samara, Russia}


\begin{abstract}
We study prompt $\Upsilon(nS)$ hadroproduction ($n=1,2,3$) invoking
the hypothesis of gluon Reggeization in $t-$channel exchanges at
high energy and the factorization formalism of nonrelativistic
quantum chromodynamics at leading order in the strong-coupling
constant $\alpha_s$ and the relative velocity $v$ of the bound quarks. The
transverse-momentum distributions of prompt $\Upsilon(nS)$-meson
production measured by the ATLAS Collaboration at the CERN LHC are
fitted to obtain the color-octet nonperturbative long-distance
matrix elements, which are used to predict prompt $\Upsilon(nS)$
production spectra measured by the CMS and LHCb Collaborations. At
the numerical calculation, we adopt the Kimber-Martin-Ryskin
prescription to derive unintegrated gluon distribution function of
the proton from its collinear counterpart, for which we use the
Martin-Roberts-Stirling-Thorne set. We find good agreement with
measurements by the ATLAS, CMS and LHCb Collaborations at the LHC at
the hadronic c.m.\ energy $\sqrt S=7$ TeV as well as with
measurements by the CDF Collaboration at the Fermilab Tevatron.
\end{abstract}

\maketitle

\section{Introduction}
\label{sec:one}

The production of charmonium and bottomonium states at hadron
colliders has provided a useful laboratory for testing the high-energy
limit of quantum chromodynamics (QCD) as well as the interplay of
perturbative and nonperturbative phenomena in QCD. The additional
interest to heavy quarkonium production is motivated by the idea
than it can be distinguished as a signal manifesting new phenomena,
such as quark-gluon plasma production, color-transparency, associated
Higgs boson production and so on. The experimental study of
bottomonium production at the Large Hadron Collider (LHC) is
included in programs of main CERN Collaborations: ATLAS
\cite{atlas},  CMS \cite{cms}, and LHCb \cite{lhcb}.

The total collision energies,  $\sqrt{S}=7$~TeV or 14~TeV at the
LHC, sufficiently exceed the characteristic scale $\mu$ of the
relevant hard processes, which is of order of quarkonium transverse
mass $M_T=\sqrt{M^2+p_T^2}$, {\it i.e.}\ we have
$\Lambda_\mathrm{QCD}\ll\mu\ll\sqrt{S}$. In this high-energy regime,
so called "Regge limit", the contribution of partonic subprocesses
involving $t$-channel parton (gluon or quark) exchanges to the
production cross section can become dominant. Then the transverse
momenta of the incoming partons and their off-shell properties can
no longer be neglected, and we deal with "Reggeized" $t$-channel
partons. These $t$-channel exchanges obey multi-Regge kinematics
(MRK), when the particles produced in the collision are strongly
separated in rapidity. If the same situation is realized with groups
of particles, then quasimulti-Regge kinematics (QMRK) is at work. In
the case of $\Upsilon(nS)-$meson inclusive production, this means
the following: $\Upsilon(nS)$-meson (MRK) or $\Upsilon(nS)$-meson
plus gluon jet (QMRK) are produced in the central region of rapidity,
while other particles are produced with large modula of rapidities.

The parton Reggeization approach \cite{QMRK,Lipatov95} is
particularly appropriate for high-energy phenomenology. We see, the
assumption of a dominant role of MRK or QMRK production mechanisms
at high energy works well. The parton Reggeization approach is based
on an effective quantum field theory implemented with the
non-Abelian gauge-invariant action including fields of Reggeized
gluons \cite{BFKL} and Reggeized quarks \cite{LipatoVyazovsky}.
Reggeized partons interact with quarks and Yang-Mills gluons in a
specific way. Recently, in Ref.\cite{Antonov}, the Feynman rules for
the effective theory of Reggeized gluons were derived for the
induced and some important effective vertices. This approach was
successfully applied to interpret the production of isolated jets
\cite{KSS2011}, dijet azimuthal decorrelations \cite{dijet2013},
prompt photons \cite{SVADISy}, diphotons \cite{SVAdiy}, charmed
mesons \cite{PRD}, bottom-flavored jets \cite{PRb}, Drell-Yan lepton
pairs \cite{NNS_DY} measured at the Fermilab Tevatron, at the DESY
HERA and at the CERN LHC, especially in the small-$p_T$ regime,
where $p_T<<\sqrt{S}$.

We suggest the MRK or QMRK production mechanisms to be dominant also for
heavy-quarkonium production at the LHC. Using the Feynman rules
\cite{Antonov} for the effective theory, we can construct heavy-quarkonium production amplitudes in the non-relativistic
QCD (NRQCD)\cite{NRQCD,Maltoni}. The factorization formalism of the
NRQCD  is a rigorous theoretical framework for the description of
heavy-quarkonium production and decay. The factorization hypothesis
of NRQCD assumes the separation of the effects of long and short
distances in heavy-quarkonium production. NRQCD is organized as a
perturbative expansion in two small parameters, the strong-coupling
constant $\alpha_s$ and the relative velocity $v$ of the heavy
quarks inside a heavy quarkonium.

Our previous analysis of charmonium \cite{KSVcharm, SVpepan} and
bottomonium \cite{KSVbottom,SVpepan} production at the Fermilab
Tevatron and charmonium production \cite{NSScharm} at the CERN LHC
using the high-energy factorization scheme and the NRQCD approach has
shown the efficiency of such type of high-energy phenomenology. In
this paper we perform calculations for the  prompt
$\Upsilon(nS)$-meson transverse momentum spectra at the CERN LHC to
obtain color-octet nonperturbative matrix
elements (NMEs) by fitting procedure using experimental data from the ATLAS Collaboration~\cite{atlas}.
Then we predict prompt $\Upsilon(nS)$-meson spectra,
which were measured recently by the  CMS~\cite{cms} and
LHCb~\cite{lhcb} CERN LHC Collaborations at the energy of $\sqrt
S=7$ TeV and a few years before by the CDF~\cite{cdf} Fermilab
Tevatron Collaboration at the energy of $\sqrt S=1.8$ TeV. We find a
good agreement of our calculations and experimental data.

\section{Model}

Working at the leading order (LO) in  $\alpha_s$ and $v$ we consider
the following partonic subprocesses, which describe bottomonium
production at high energy:
\begin{eqnarray}
R(q_1) + R(q_2) &\to& {\cal H}
[{^3P}_J^{(1)},{^3S}_1^{(8)},{^1S}_0^{(8)},{^3P}_J^{(8)}](p),
\label{eq:RRtoH}\\
 R(q_1) + R(q_2) &\to& {\cal H} [{^3S}_1^{(1)}](p) + g(p'),
\label{eq:RRtoHG}
\end{eqnarray}
where $R$ is a Reggeeized gluon and $g$ is an on-shell Yang-Mills gluon,
respectively, with four-momenta indicated in parentheses, ${\cal
H}[n]$ is a physical bottomonium state, $n={}^{2S+1}L_J^{(1,8)}$
is a $b\bar b$ Fock state with a spin $S$, total
angular momentum $J$, orbital angular momentum $L$ and with the
color-singlet $(1)$ or the color-octet $(8)$ quantum numbers.

In the general case, the partonic cross section of bottomonium
production receives from the $b\bar b$ Fock state
$[n]=[{}^{2S+1}L_J^{(1,8)}]$ the contribution \cite{NRQCD,Maltoni}
\begin{equation}
d\hat \sigma (R + R \to b\bar b[{}^{2S+1}L_J^{(1,8)}] \to {\cal
H})=d\hat \sigma (R + R \to b\bar
b[^{2S+1}L_J^{(1,8)}])\frac{\langle {\cal O}^{\cal
H}[^{2S+1}L_J^{(1,8)}]\rangle}{N_\mathrm{col}N_\mathrm{pol}},
\end{equation}
where $N_\mathrm{col}=2 N_c$ for the color-singlet state,
$N_\mathrm{col}=N_c^2-1$ for the color-octet state, and
$N_\mathrm{pol}=2J+1$, $\langle {\cal O}^{\cal
H}[^{2S+1}L_J^{(1,8)}]\rangle$ are the NMEs. They satisfy the
multiplicity relations
\begin{eqnarray}
\langle{\cal O}^{\psi(nS)}[^3P_J^{(8)}]\rangle&=&(2J+1)\langle{\cal
O}^{\psi(nS)}[^3P_0^{(8)}]\rangle,\nonumber\\
\langle{\cal O}^{\chi_{cJ}}[^3P_J^{(1)}]\rangle&=&(2J+1)\langle{\cal
O}^{\chi_{c0}}[^3P_0^{(1)}]\rangle,\nonumber\\
\langle{\cal O}^{\chi_{cJ}}[^3S_1^{(8)}]\rangle&=&(2J+1)\langle{\cal
O}^{\chi_{c0}}[^3S_1^{(8)}]\rangle,
\end{eqnarray}
which follow from heavy-quark spin symmetry in the LO in $v$. The color-singlet NMEs can be obtained from values of
quarkonium radial wave function and its derivative in the origin by the following formulas:
\begin{eqnarray}
\left\langle {\cal O}^{{\cal H}_J} \left[^3S_1^{(1)}\right]\right\rangle = 2N_c(2J+1)\frac{1}{4\pi} |R(0)|^2,\label{CSNME1}\\
\left\langle {\cal O}^{{\cal H}_J} \left[^3P_J^{(1)}\right]\right\rangle = 2N_c(2J+1)\frac{3}{4\pi} |R'(0)|^2.\label{CSNME2}
\end{eqnarray}
The partonic cross section of $b\bar b$ production is defined as
\begin{equation}
d\hat\sigma(R + R \to b\bar
b[^{2S+1}L_J^{(1,8)}])=\frac{1}{I}\overline{|{\cal A}(R + R \to
b\bar b[^{2S+1}L_J^{(1,8)}])|^2}d\Phi,\label{eq:dsigma}
\end{equation}
where $I=2 x_1 x_2 S$ is the flux factor of the incoming particles,
which is taken as in the collinear parton model \cite{KTCollins},
${\cal A}(R + R \to b\bar b[{}^{2S+1}L_J^{(1,8)}])$ is the
production amplitude, the overbar indicates average (summation) over
initial-state (final-state) spins and colors, and $d\Phi$ is the
invariant phase space volume of the outgoing particles. This
convention implies that the cross section in the high-energy
factorization scheme is normalized approximately to the cross
section for on-shell gluons in the collinear parton model when ${\bf
q}_{1T}={\bf q}_{2T}={\bf 0}$.

Earlier we have found the LO results for the squared amplitudes of
subprocesses~(\ref{eq:RRtoH}) and (\ref{eq:RRtoHG}) using the Feynman rules of Ref.~\cite{Antonov}.  The formulas
for the squared amplitudes $\overline{|{\cal A}(R + R \to b\bar
b[^{2S+1}L_J^{(1,8)}])|^2}$ for the $2\to 1$ subprocesses
(\ref{eq:RRtoH}) are listed in Eq. (27) of Ref.~\cite{KSVcharm}. The
analytical result in case of the $2\to 2$ subprocess
(\ref{eq:RRtoHG}) is presented in Ref.~\cite{SVpepan}.

Exploiting the hypothesis of high-energy factorization, we may write
the hadronic cross section $d\sigma$ as a convolution of partonic
cross section $d\hat \sigma$ with unintegrated parton distribution
functions (PDFs) $\Phi_g^p(x,t,\mu^2)$ of Reggeized gluon in the
proton, as
\begin{eqnarray}
d\sigma(p + p \to {\cal H} + X)&=&  \int\frac{d x_1}{x_1}
\int\frac{d^2{\bf q}_{1T}}{\pi} \Phi_g^p\left(x_1,t_1,\mu^2\right)
\int\frac{d x_2}{x_2} \int\frac{d^2 {\bf q}_{2T}}{\pi}
\nonumber\\
&&{}\times\Phi_g^p\left(x_2,t_2,\mu^2\right) d\hat\sigma(R + R \to
{\cal H}+X). \label{eq:KT}
\end{eqnarray}
$t_1=|{\bf q}_{1T}|^2$, $t_2=|{\bf q}_{2T}|^2$, $x_1$ and $x_2$
are the fractions of the proton momenta passed on to the Reggeized
gluons, and the factorization scale $\mu$ is chosen to be of order
$M_T$. The collinear and unintegrated gluon distribution functions
are formally related as
\begin{equation}
xG^p(x,\mu^2)=\int^{\mu^2} \Phi_g^p(x,t,\mu^2)dt, \label{eq:xG}
\end{equation}
so that, for ${\bf q}_{1T}={\bf q}_{2T}={\bf 0}$, we recover the
conventional factorization formula of the collinear parton model,
\begin{equation}
d\sigma(p + p \to {\cal H}+\!X)=\int{d x_1}G^p(x_1,\mu^2) \int{d
x_2} G^p(x_2,\mu^2) d\hat \sigma(g + g \to {\cal H} + X).
\label{eq:PM}
\end{equation}

We now describe how to evaluate the differential hadronic cross
section from Eq.~(\ref{eq:KT}) combined with the squared amplitudes
of the $2\to 1$ and $2\to 2$ subprocesses~(\ref{eq:RRtoH}) and
(\ref{eq:RRtoHG}), respectively. The rapidity and pseudorapidity of
a bottomonium state with four-momentum $p^\mu=(p^0,{\bf p}_{T},p^3)$
are defined as follows
\begin{equation}
y=\frac{1}{2}\ln\frac{p^0+p^3}{p^0-p^3},\quad
\eta=\frac{1}{2}\ln\frac{|{\bf p}|+p^3}{|{\bf p}|-p^3}.
\end{equation}
In the following, we shall also use the shorthand notation $p_T=|{\bf p}_{T}|$ etc.
for the absolute of the transverse two-momentum.


 The master formula for the $2\to1$
subprocess~(\ref{eq:RRtoH}) can be presented by the following way:
\begin{eqnarray}
&&\frac{d\sigma(p + p \to {\cal H} + X)} {d p_Td y} = \frac{p_T}{(
p_T^2+M^2)^2} \int{dt_1}\int{d \varphi_1}
\nonumber\\
&&{}\times \Phi_g^p(\xi_1,t_1,\mu^2) \Phi_g^p(\xi_2,t_2,\mu^2)
\overline{|{\cal A}(R + R \to {\cal H})|^2},
\end{eqnarray}
where $t_2=t_1+p_T^2-2 p_T \sqrt{t_1}\cos (\phi_1)$,
$\xi_1=(p^0+p^3)/\sqrt{S}$, $\xi_2=(p^0-p^3)/\sqrt{S}$ and the
relation $\xi_1\xi_2 S=M_T^2=p_T^2+M^2$ has been taken into account.

Than, we write the master formula for the $2\to2$
subprocess~(\ref{eq:RRtoHG}):
\begin{eqnarray}
&&\frac{d\sigma(p + p \to {\cal H} + X)}{d{p_T d y}}=\frac{p_T}{(2
\pi)^3} \int{dt_1}\int{d \varphi_1} \int{d x_2}\int{dt_2}\int{d
\varphi_2}\nonumber\\&&{} \times \Phi_g^p(x_1,t_1,\mu^2)
\Phi_{g}^p(x_2,t_2,\mu^2) \frac{\overline{|{\cal A}(R + R \to {\cal
H} + g)|^2}}{ (x_2 - \xi_2)(2 x_1 x_2 S)^2},
\end{eqnarray}
where $\phi_{1,2}$ are the angles enclosed between ${\bf
q}_{1,2T}$ and the transverse momentum ${\bf p}_T$ of ${\cal
H}$,
\begin{equation}
x_1=\frac{1}{(x_2 - \xi_2) S}\left[({\bf q}_{1T}+{\bf q}_{2T} - {\bf
p}_T)^2- M^2 - |{\bf p}_{T}|^2 + x_2 \xi_1 S\right].
\end{equation}


In our numerical analysis, we adopt as our default the prescription
proposed by Kimber, Martin, and Ryskin (KMR) \cite{KMR} to obtain
unintegrated gluon PDF of the proton from the conventional
integrated one, as implemented in Watt's code \cite{Watt}.   As
input for these procedures, we use the LO set of the
Martin-Roberts-Stirling-Thorne (MRST) \cite{MRST} proton PDF as our
default.  Throughout our analysis the renormalization and
factorization scales are identified and chosen to be $\mu=\xi M_T$,
where $\xi$ is varied between 1/2 and 2 about its default value 1 to
estimate the theoretical uncertainty due to the freedom in the
choice of scales. The resulting errors are indicated as shaded bands
in the figures.

\section{Results}

First of all, to extract the color-octet NMEs of the $\Upsilon(nS)$-mesons, we
perform a fit to the ATLAS Collaboration data~\cite{atlas}
on prompt $\Upsilon(nS)$-meson production collected in proton-proton collisions at the
energy $\sqrt{S}=7$~TeV in the two regions of rapidity  $|y|<1.2$
and $1.2<|y|<2.25$. This data set has the smallest
statistical uncertainties and covers the largest interval in transverse momentum, namely $0<p_T<70$~GeV,
in comparison with data sets from other CERN LHC
collaborations. The
$\Upsilon(3S)$ mesons are produced only directly via color-singlet
and color-octet production mechanisms, $\Upsilon(1S)$ and
$\Upsilon(2S)$ mesons are produced promptly, i.e. directly or through nonforbidden decays
of higher-lying $\chi_{bJ}$ and $\Upsilon(nS)$ mesons, including cascade transitions
such as $\Upsilon(3S)\to\chi_{b1}\to \Upsilon(1S)$. Notice that the contributions to prompt $\Upsilon(1S)$ and $\Upsilon(2S)$
production due to a feed-down are non-negligible for cascade decays up to the third order only.
Thus, we introduce the matrix $\hat{B}$ composed from the corresponding branching ratios
 of feed-down decays, which are extracted of the experimental data~\cite{PDG}.
 Then, the feed-down contribution can be evaluated as a product of the column-vector
  of direct cross sections and the matrix\ $ \hat{B}+\hat{B}^2+\hat{B}^3$.  For the reader's convenience,
  we list the matrix $\hat{B}$ in the Table~\ref{TableI}. Since the $\Upsilon(nS)$ mesons are identified
  through their decays to $\mu^+ \mu^-$ pairs, we have to include the corresponding branching fractions,
  which we adopt from the recent
    Particle Data Group (PDG) report~\cite{PDG}, $B\left(\Upsilon(1S)\to \mu^+\mu^-\right)=0.0248$, $B\left(\Upsilon(2S)\to \mu^+\mu^-\right)=0.0193$,
 and  $B\left(\Upsilon(3S)\to \mu^+\mu^-\right)=0.0218$.

Our fits of the color-octet NMEs include six experimental data samples,
which come as $p_T$ distributions of $\Upsilon(nS)$ mesons prompt
production. In
Table~\ref{TableII} we list our fit results, along with the values of color-singlet NMEs
used. The last ones we determine by the equations
(\ref{CSNME1}, \ref{CSNME2}) using the quarkonium wave
functions and their derivatives evaluated at the origin from
potential models, see Ref.~\cite{PotMod}.

 We perform a fit procedure with the positivity constraint on color-octet NMEs. Also, turning to the previous studies of charmonium and
bottomonium production at the Fermilab Tevatron, we assume
that $\chi_{bJ}$ mesons are produced directly only via the
color-singlet mechanism, so we put the corresponding color-octet matrix elements equal to zero, see Table~\ref{TableII}.

 The errors on the fit results are determined by varying in turn each NME up and down about its central value until the value of $\chi^2$ is increased
 by unity keeping all other NMEs fixed at their central values.
 We found a quantity $\chi^2/d.o.f.$ to have a quite large value of $29.9$. However, this result is foreseen,
 because in spite of the very small statistical and systematical errors of the ATLAS
  data, which were used for the $\chi^2$-procedure and indicated in the figures,
  in the region of $p_T<10$~GeV these data contain a huge uncertainty due to polarization effects~\cite{atlas}. As the $\Upsilon(nS)$
  production cross section is extracted from the inclusive $\mu^+ \mu^-$ production cross section with the certain kinematical cuts,
   in the region of small $p_T$ the result of such extraction is significantly dependent on the assumptions on polarization of produced quarkonium.
   The inclusion of this big uncertainty makes a fit too insensitive to the value of the cross section in the low-$p_T$ region, so we decided
   to skip this uncertainty and increase the values of NMEs errors $\sqrt{\chi^2/d.o.f.}$ times, like it is implemented by the PDG
   in the cases of large values of $\chi^2/d.o.f.$~\cite{PDG}.
  As it can be seen from Table~\ref{TableII}, the fit procedure strongly suppresses all color-octet NMEs except the $^3S_1^{(8)}$ NMEs
   for $\Upsilon(nS)$ mesons.

In Figs.~\ref{figatlas}--\ref{figfrac}, we compare our
predictions obtained in the LO NRQCD and the parton Reggeization
approach with the data on $\Upsilon(nS)$ mesons prompt
production, measured by the ATLAS Collaboration
(Fig.~\ref{figatlas}), by the CMS Collaboration (Fig.~\ref{figcms}),  by
the LHCb Collaboration (Figs.~\ref{figlhcb1}--\ref{figfrac}), and by the CDF
Collaboration (Fig.~\ref{figcdf}). It is important, the
experimental data \cite{atlas,cms,lhcb,cdf} depend on the assumption
of polarization of produced $\Upsilon(nS)$ mesons slightly. We
perform our calculations and make a comparison to the data in a case of
non-polarized $\Upsilon(nS)$ meson production.

Let us mention general features of $\Upsilon(nS)$ meson $p_T$
distributions which are evident under all considered experimental
conditions from LHC and Tevatron Colliders. The production of
$\Upsilon(1S)$ and $\Upsilon(2S)$ mesons for $p_T<20$ GeV and for
all rapidities is dominated by the color-singlet mechanism while the
color-octet production mechanism dominates only at large transverse
momenta $p_T\geq 20$~GeV. This result confirms a naive estimation
that at large transverse momentum the gluon fragmentation to
bottomonium via gluon splitting to $b\bar b$ pair in the
$^3S_1^{(8)}$ color-octet state should be more important. Roughly
speaking, $\Upsilon(1S)$ prompt production can be described using
color-singlet production mechanism only. It follows from the fact
that a significant part of $\Upsilon(1S)$ mesons is produced
directly through the color singlet mechanism
 or via cascade decays of higher-lying $P$-wave states $\chi_{Jb}$.

  The more important role of the color-octet mechanism appears in the production of $\Upsilon(3S)$ mesons.
  For the rapidities $|y|<3$ its contribution amounts more than 50\% of the cross section already from $p_T\geq 12$~GeV.
   This boundary rapidly decreases with growth of rapidity, and, as it follows from the comparison with LHCb data in Fig.~\ref{figlhcb1},
    for the rapidities $|y|>3$ the color-octet mechanism dominates at all values of $p_T$. In the region of large rapidities $|y|>3.5$
     shown in the Fig.~\ref{figlhcb2}, our model tends to overestimate the experimental cross section. The same feature was also observed
     in our study of charmonium production~\cite{NSScharm}, and probably it reflects a violation of the QMRK condition at large values of $|y|$.

The $P$-wave bottomonium production in the parton Reggeization model can be
described well in the color-singlet model, see Table~I, as in the case of the $P$-wave charmonium production
\cite{Teryaev,KSVcharm}. In the Fig.~
\ref{figfrac}, our prediction on the fraction of $\Upsilon(1S)$
mesons produced in the decays of $\chi_{bJ}(1P)$ mesons is compared
with LHCb data \cite{lhcbfr}. We find a good agreement of this data
with our prediction, despite the fact that this dataset was not
included to the fit procedure. We predict additionally the fraction
of $\Upsilon(2S)$ mesons produced in the decays of $\chi_{bJ}(2P)$
mesons.

 In Fig.~\ref{figcdf} one can find the CDF data~\cite{cdf} on prompt $\Upsilon(nS)$ production at $|y|<0.4$ and $\sqrt{S}=1.8$~TeV to be
 also well described by our model with the values of color-octet NMEs from Table~II.

Comparing our results with the recent studies of $\Upsilon(nS)$
meson hadroproduction in the conventional collinear parton model
performed in  LO~\cite{bbloCO} and  full NLO approximation of NRQCD
formalism \cite{bbnloCO} or in the non-complete NNLO$^{*}$
approximation of color-singlet model \cite{bbnloCS}, we should
emphasize the following. Oppositely to the calculations in the
collinear parton model~\cite{bbloCO,bbnloCO,bbnloCS}, we describe
data at the small transverse momenta of $\Upsilon(nS)$ mesons
(especially $\Upsilon(1S)$ meson) well, down to $p_T=0$, at all
values of rapidity. The LO heavy quarkonium production amplitudes in
the parton Reggeization approach are finite at the $p_T=0$ as well
as the unintegrated gluon PDFs. Our predictions at the small
$p_T<10$~GeV have relatively large uncertainties, about factor 2,
related with the strong dependence of cross section on the choice of
the factorization scale $\mu$. However, the central line
corresponding to the default choice of $\mu=M_T$, lies rather close
to the average values of the experimental data. The inclusion of
small $p_T$ region in a fit range is very important to separate
different color-octet contributions and it changes the relative
values of color-octet NMEs, comparing to the case when only large
$p_T$ region is taken into account.

  The region of small $p_T$ is also important to test the possibility of negative color-octet NMEs.
   This possibility was firstly supposed in~\cite{KniehlNEG}, and it is used in modern full NLO studies of bottomonium production,
   such as Ref.~\cite{bbnloCO}.
  To perform the fit of color-octet NMEs without the positivity condition (unconstrained fit), it is necessary to include into the fit
   as much experimental data as possible. The most important constraints on the values of color-octet NMEs are coming from the data on
   $\Upsilon(1S)$ fraction from the $\chi_{bJ}(1P)$ decays~\cite{lhcbfr}, and from the LHCb data at rapidities $|y|>3$~\cite{lhcb}. In Ref.~\cite{bbnloCO},
    also the data on $\Upsilon(1S)$ polarization were taken into account.

  In our model, the unconstrained fit of the ATLAS data together with data on $\Upsilon(1S)$ fraction from the $\chi_{bJ}(1P)$ decays,
  significantly improves the description of experimental data at small rapidities, but leads to negative values of the cross section
  of $\Upsilon(3S)$ production for $p_T<8$ GeV and $|y|>3$. To avoid this problem, it is necessary to include LHCb data to the fit, which
  greatly suppresses the negative values of color-octet NMEs. So, we conclude that negative values of fit parameters
   are not necessary for the description of data at small values of $p_T$ and large rapidities. Moreover, the positivity condition
   improves the predictive power of the model, allowing a reasonably good description of all present data on cross sections with
   just three free parameters.

Recently, the analysis of prompt $\Upsilon(nS)$ production at the
LHC in view of the $k_T-$factorization approach was considered in
Ref.~\cite{baranovKT}. Oppositely to our conclusions, it was found
that data from CMS~\cite{cms} and LHCb~\cite{lhcb} Collaborations
can be described using the color-singlet production mechanism only.
Let us discuss here the main differences between our model and the model of Ref.~\cite{baranovKT}.

In Ref.~\cite{baranovKT}, it is suggested the occurrence of additional
feed-down contribution from $\chi_{bJ}(3P)$ mesons, which is absent
in our calculation. The first measurement of $\chi_{bJ}(3P)$
bottomonium state estimates its mass as $m(3P)=10.530\pm 0.014$~GeV~
\cite{chi3P}. This value is very close to the mass-threshold
of open $b$-quark production, $m_{thr}= 2m(B^\pm)\simeq 10.558$~GeV,
and $\chi_{bJ}(3P)$ meson seems as a very unstable state with
unknown branching ratio $B(\chi_{bJ}(3P)\to \Upsilon(3S))$. We
guess the inclusion of this contribution is still in dispute, moreover,
it is non-negligible for the $\Upsilon(3S)$ meson spectra only.

Comparing the relative contributions of direct and feed-down
production mechanisms, which are presented in Ref.~\cite{baranovKT},
we see that a direct contribution dominates at small $p_T$ and a
feed-down contribution dominates at large values of $p_T$ exceeding
$5-7$~GeV. This result contradicts to the recent measurements of the
LHCb Collaboration \cite{lhcbfr}, which demonstrate that a fraction
of $\Upsilon(1S)$ mesons from decays of $\chi_{bJ}(1P)$ mesons is
approximately constant at all values of $p_T$ (see Fig.
\ref{figfrac}), it is about $30$~\%. Such a way, in
Ref.~\cite{baranovKT} the value of feed-down contribution to prompt
$\Upsilon(nS)$ production is strongly overestimated, and in fact we
need to take into account a color-octet production mechanism to
describe the experimental data \cite{atlas,cms,lhcb}. For precise
comparison with results obtained in Ref.~\cite{baranovKT}, we show
in Fig.~\ref{figsinglet} our prediction for color-singlet
contributions only. We see, the additional contribution is needed
already at the $p_T\geq 10$~GeV and this contribution should have a
gently sloping of the transverse momentum spectrum comparing to
color-singlet contribution, that belongs only to the contribution of
color-octet $^3S_1^{(8)}$ state, see Fig.~1 in Ref.~\cite{KSVcharm}.

We compare the results of our fit for color-octet NMEs with the values
recently obtained in full NLO calculation of NRQCD
approach~\cite{bbnloCO}. If we take into account some differences
between fit procedures used here and in Ref.~\cite{bbnloCO} and
perform a fit in a way of Ref.~\cite{bbnloCO}, we obtain
the very similar values of color-octet NMEs for $^3S_1^{(8)}$ states.
Such an agreement demonstrates a validity of factorization
hypothesis in the bottomonium production in hadronic collisions,
i.e. an independence of the $b\bar b$ production mechanism from the
nonperturbative bottomonium formation at the last step. It is
necessary to note, that a same consent between LO results obtained in
the parton Reggeization approach and NLO results obtained in the
collinear parton model is also observed describing charmonium
production processes, see Refs.~\cite{KSVcharm,SVpepan,NSScharm}.

The present study along with the previous investigations in the
parton Reggeization approach
\cite{KSS2011,dijet2013,SVADISy,SVAdiy,PRD,PRb,NNS_DY,KSVcharm,SVpepan,KSVbottom,NSScharm}
demonstrates the {important} role of (quasi)multi-Regge kinematics in
particle production at high energies, this feature is out of account
in the collinear parton model. {Such a way, we find the approach
based on the effective theory of Reggeized partons
\cite{BFKL,Lipatov95} and high-energy factorization scheme with
unintegrated PDFs, in which the large logarithmic terms
($\ln(\mu^2/\Lambda_{QCD}^2)$, and $\ln(S/\mu^2)$) are resummed in
all orders of strong coupling constant $\alpha_s$, to be more
adequate for the description of experimental data than fixed order
calculations in $\alpha_s$ in the framework of collinear parton
model.

\section{Conclusions}
\label{sec:five}

The CERN LHC is currently probing particle physics at terascale
c.m.\ energies $\sqrt{S}$, so that the hierarchy
$\Lambda_\mathrm{QCD}\ll\mu\ll\sqrt{S}$, which defines the MRK and
QMRK regimes, is satisfied for processes of heavy quark and heavy
quarkonium production in the central region of rapidity, where $\mu$
is of order of their transverse mass. In this paper, we studied  QCD
processes of particular interest, namely prompt $\Upsilon(nS)$
hadroproduction, at LOs in the parton Reggeization approach and
NRQCD approach, in which they are mediated by $2\to1$ and $2\to 2$
partonic subprocesses initiated by Reggeized gluon collisions.

We found by the fit of ATLAS Collaboration data~\cite{atlas}
the numerical values of the color-octet NMEs. Using these NMEs, we
nicely describe recent LHC and old Tevatron data for prompt
$\Upsilon(nS)$ meson production measured by ATLAS~\cite{atlas},
CMS~\cite{cms} and LHCb~\cite{lhcb} Collaborations at the whole
presented range of $\Upsilon(nS)$ transverse momenta and rapidity
$y$.  Here and in
Refs.~\cite{KSS2011,dijet2013,SVADISy,SVAdiy,PRD,PRb,NNS_DY,KSVcharm,SVpepan,KSVbottom,NSScharm},
the parton Reggeization approach was demonstrated to be a powerful tool
for the theoretical description of QCD processes in the high-energy
limit.

\section{Acknowledgements}

We are grateful to  B.~A.~Kniehl and N.~N.~Nikolaev for useful
discussions. The work of V.~S. was supported by the Ministry for
Science and Education of the Russian Federation under Contract
No.~2.870.2011. The work of  M.~N. and A.~S. was supported in part
by the Russian Foundation for Basic Research under Grant
12-02-31701-mol-a. The work of M.~N. is supported also by the Grant
of the Graduate Students Stipend Program of the Dynasty Foundation.

\newpage

\begin{table}
 \caption{\label{TableI}Matrix of the inclusive branching fractions. All known branchings are taken from~\cite{PDG},
  others are supposed to be equal to zero.}
\begin{ruledtabular}
  \begin{tabular}{c|ccccccccc}
  Out $\setminus$ In & $\Upsilon(3S)$ & $\chi_{b2}(2P)$ & $\chi_{b1}(2P)$& $\chi_{b0}(2P)$ &$\Upsilon (2S)$&
  $\chi_{b2}(1P)$ & $\chi_{b1}(1P)$& $\chi_{b0}(1P)$  \\
\hline

$\chi_{b2}(2P)$&0.131 &--&--&--&--&--&--&--\\

$\chi_{b1}(2P)$&0.126 &--&--&--&--&--&--&--\\

$\chi_{b0}(2P)$&0.059 &--&--&--&--&--&--&--\\

$\Upsilon(2S)$&0.106 &0.106 &0.199 &0.046 &--&--&--&--\\

$\chi_{b2}(1P)$&0.0099 &0.0051 &--&--&0.0715 &--&--&--\\

$\chi_{b1}(1P)$&0.0009 &--&0.0091 &--&0.069 &--&--&--\\

$\chi_{b0}(1P)$&0.0027 &--&--&--&0.0038 &--&--&--\\

$\Upsilon(1S)$&0.0657 &0.081 &0.108 &0.009 &0.2652 &0.191 &0.339 &0.0176 \\
  \end{tabular}
\end{ruledtabular}
\end{table}

 \begin{table}
 \caption{\label{TableII} The color-singlet and color-octet NMEs used in the calculation.}
 \begin{ruledtabular}
 \begin{tabular}{c|c}
  NME & Fit in LO of parton Reggeization approach. \\
  \hline
  $\left\langle {\cal O}^{\Upsilon(1S)}\left[^3S_1^{(1)}\right]\right\rangle\times$ GeV$^{-3}$ & $9.28$\\
  $\left\langle {\cal O}^{\Upsilon(1S)}\left[^3S_1^{(8)}\right]\right\rangle\times 10^2$ GeV$^{-3}$ & $2.31\pm 0.25$\\
  $\left\langle {\cal O}^{\Upsilon(1S)}\left[^1S_0^{(8)}\right]\right\rangle\times 10^2$ GeV$^{-3}$ & $0.0\pm 0.05$ \\
  $\left\langle {\cal O}^{\Upsilon(1S)}\left[^3P_0^{(8)}\right]\right\rangle\times 10^2$ GeV$^{-5}$ & $0.0\pm 0.38$ \\
 \hline
  $\left\langle {\cal O}^{\Upsilon(2S)}\left[^3S_1^{(1)}\right]\right\rangle\times$ GeV$^{-3}$ & $4.62$ \\
  $\left\langle {\cal O}^{\Upsilon(2S)}\left[^3S_1^{(8)}\right]\right\rangle\times 10^2$ GeV$^{-3}$ & $1.51\pm 0.17$ \\
  $\left\langle {\cal O}^{\Upsilon(2S)}\left[^1S_0^{(8)}\right]\right\rangle\times 10^2$ GeV$^{-3}$ & $0.0\pm 0.01$ \\
   $\left\langle {\cal O}^{\Upsilon(2S)}\left[^3P_0^{(8)}\right]\right\rangle\times 10^2$ GeV$^{-5}$ & $0.0\pm 0.03$ \\
  \hline
  $\left\langle {\cal O}^{\Upsilon(3S)}\left[^3S_1^{(1)}\right]\right\rangle\times$ GeV$^{-3}$ & $3.54$\\
  $\left\langle {\cal O}^{\Upsilon(3S)}\left[^3S_1^{(8)}\right]\right\rangle\times 10^2$ GeV$^{-3}$ & $1.24\pm 0.13$\\
  $\left\langle {\cal O}^{\Upsilon(3S)}\left[^1S_0^{(8)}\right]\right\rangle\times 10^2$ GeV$^{-3}$ & $0.0\pm 0.01$ \\
   $\left\langle {\cal O}^{\Upsilon(3S)}\left[^3P_0^{(8)}\right]\right\rangle\times 10^2$ GeV$^{-5}$ & $0.0\pm 0.02$\\
 \hline
  $\left\langle {\cal O}^{\chi(1P)}\left[^3P_0^{(1)}\right]\right\rangle\times$ GeV$^{-5}$ & $2.03$ \\
  $\left\langle {\cal O}^{\chi(1P)}\left[^3S_1^{(8)}\right]\right\rangle\times 10^2$ GeV$^{-3}$ & $0.0$ \\
 \hline
 $\left\langle {\cal O}^{\chi(2P)}\left[^3P_0^{(1)}\right]\right\rangle\times$ GeV$^{-5}$ & $2.36$ \\
  $\left\langle {\cal O}^{\chi(2P)}\left[^3S_1^{(8)}\right]\right\rangle\times 10^2$ GeV$^{-3}$ & $0.0$ \\
 \end{tabular}
 \end{ruledtabular}
 \end{table}

\begin{figure}[ph]
\begin{center}
\includegraphics[width=1.0\textwidth, clip=]{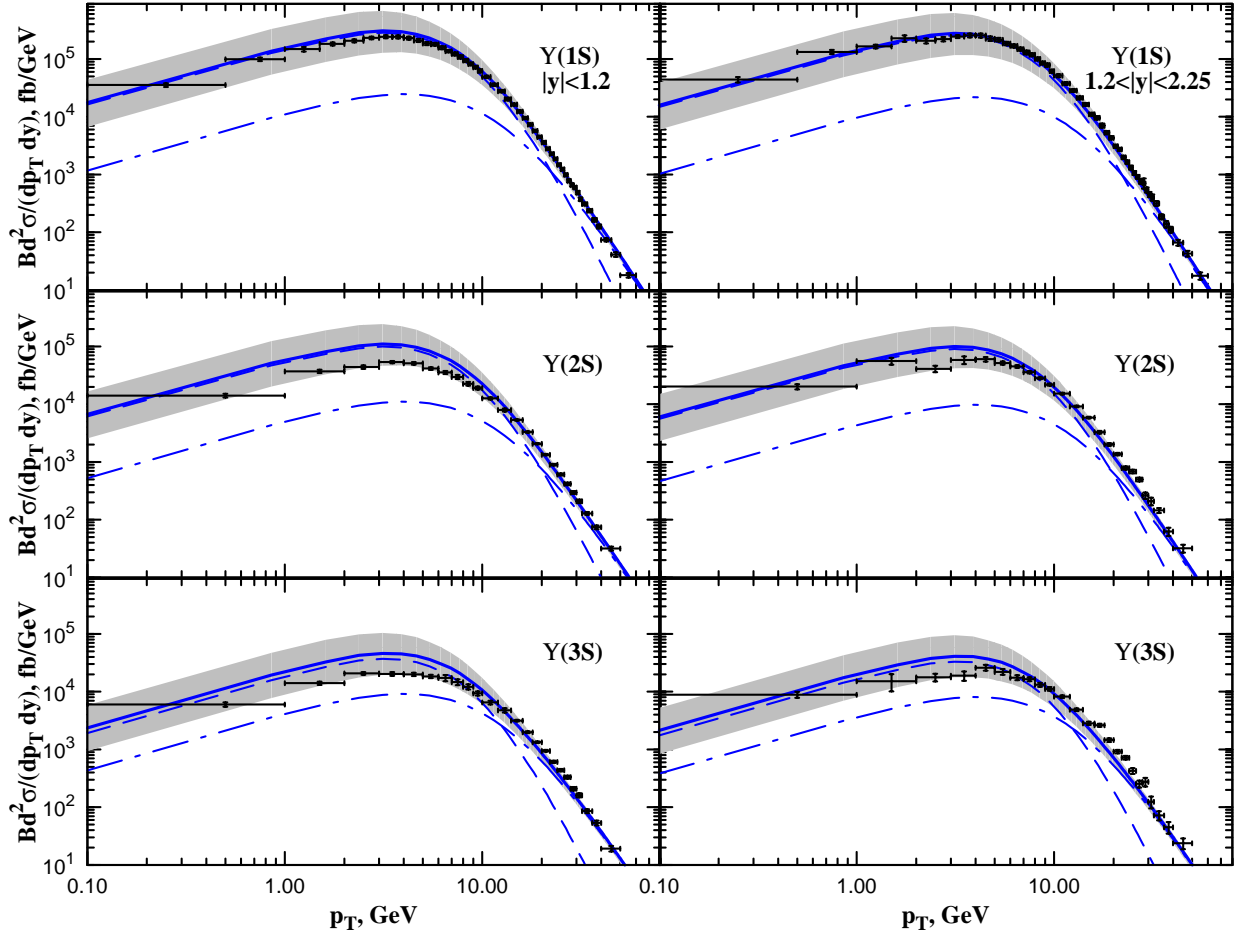}
\end{center}
\caption{ Transverse momentum distributions of prompt
$\Upsilon(1S)$, $\Upsilon(2S)$, and $\Upsilon(3S)$  hadroproduction
in $pp$ scattering with $\sqrt{S}=7$ TeV and $|y|<1.2$ (left panel)
and $1.2<|y|<2.25$ (right panel), including the respective decay
branching fractions $B(\Upsilon(nS)\to\mu^+\mu^-)$. The data are
from the ATLAS Collaboration \cite{atlas}. The curves correspond to
LO of NRQCD and parton Reggeization approach: dashed line is the
color-siglet contribution, dash-dotted line is the color-octet
contribution, solid line is their sum.}\label{figatlas}
\end{figure}

\begin{figure}[ph]
\begin{center}
\includegraphics[width=1.0\textwidth, clip=]{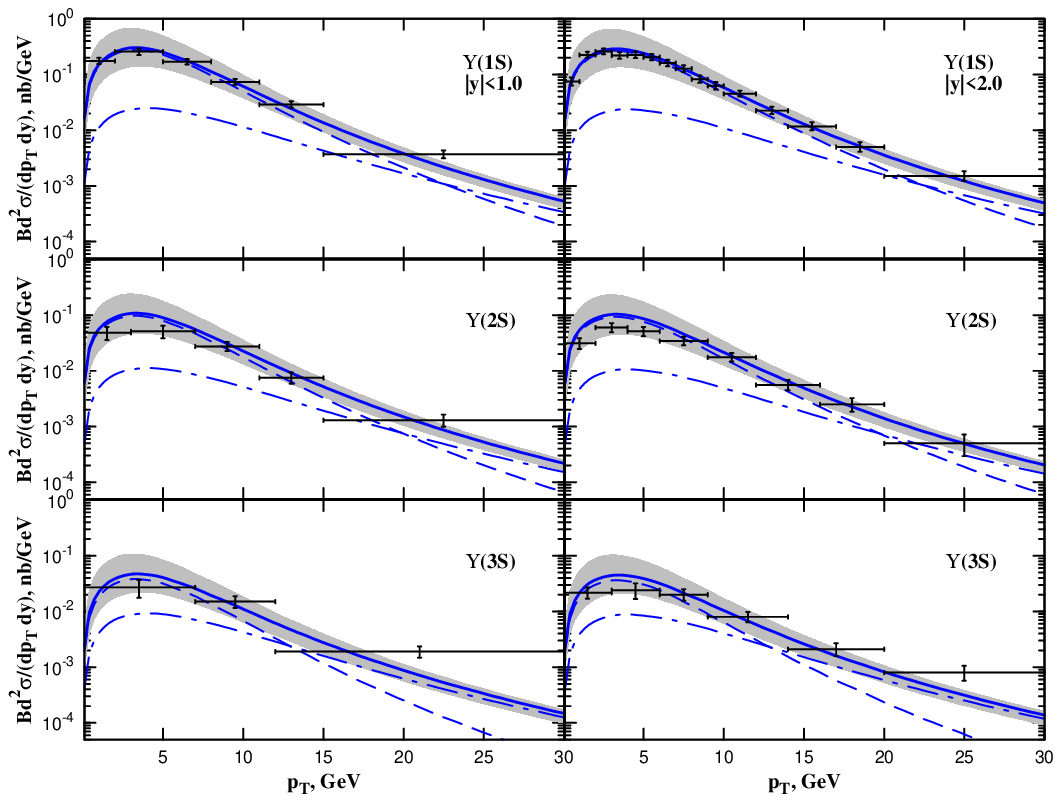}
\end{center}
\caption{ Transverse momentum distributions of prompt
$\Upsilon(1S)$, $\Upsilon(2S)$, and $\Upsilon(3S)$  hadroproduction
in $pp$ scattering with $\sqrt{S}=7$ TeV and $|y|<1.0$ (left panel)
and $|y|<2.0$ (right panel), including the respective decay
branching fractions $B(\Upsilon(nS)\to\mu^+\mu^-)$. The data are
from the CMS Collaboration \cite{cms}. The curves are the same as
in the Fig.~\ref{figatlas}.}\label{figcms}
\end{figure}

\begin{figure}[ph]
\begin{center}
\includegraphics[width=1.0\textwidth, clip=]{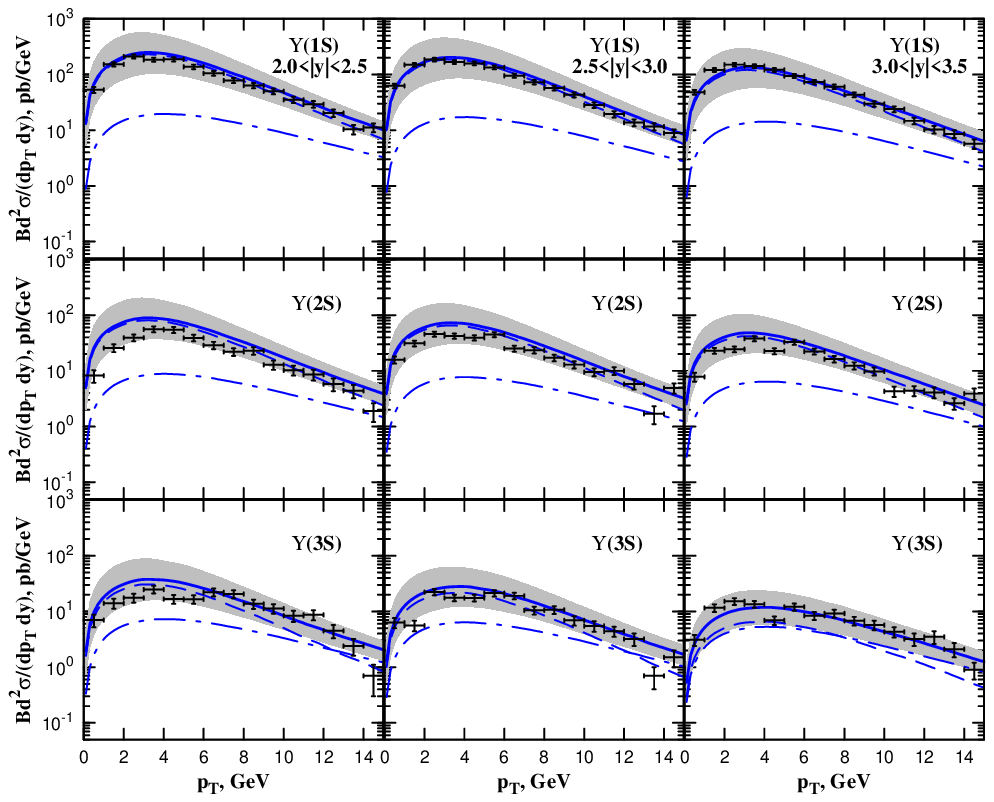}
\end{center}
\caption{ Transverse momentum distributions of prompt
$\Upsilon(1S)$, $\Upsilon(2S)$, and $\Upsilon(3S)$  hadroproduction
in $pp$ scattering with $\sqrt{S}=7$ TeV and $2.0<|y|<2.5$ (left
panel), $2.5<|y|<3.0$ (central panel), and $3.0<|y|<3.5$ (right
panel), including the respective decay branching fractions
$B(\Upsilon(nS)\to\mu^+\mu^-)$. The data are from the LHCb
Collaboration \cite{lhcb}. The curves are the same as
in the Fig.~\ref{figatlas}.}\label{figlhcb1}
\end{figure}

\begin{figure}[ph]
\begin{center}
\includegraphics[width=1.0\textwidth, clip=]{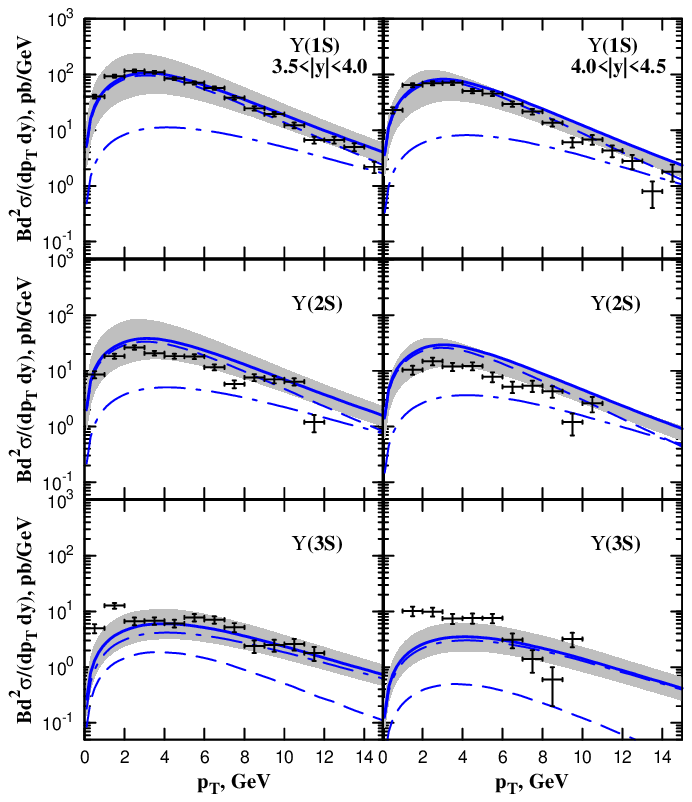}
\end{center}
\caption{ Transverse momentum distributions of prompt
$\Upsilon(1S)$, $\Upsilon(2S)$, and $\Upsilon(3S)$  hadroproduction
in $pp$ scattering with $\sqrt{S}=7$ TeV and $3.5<|y|<4.0$ (left
panel), and $4.0<|y|<4.5$ (right panel), including the respective
decay branching fractions $B(\Upsilon(nS)\to\mu^+\mu^-)$. The data
are from the LHCb Collaboration \cite{lhcb}. The curves are the same as in the
Fig.~\ref{figatlas}.}\label{figlhcb2}
\end{figure}

\begin{figure}[ph]
\begin{center}
\includegraphics[width=1.0\textwidth, clip=]{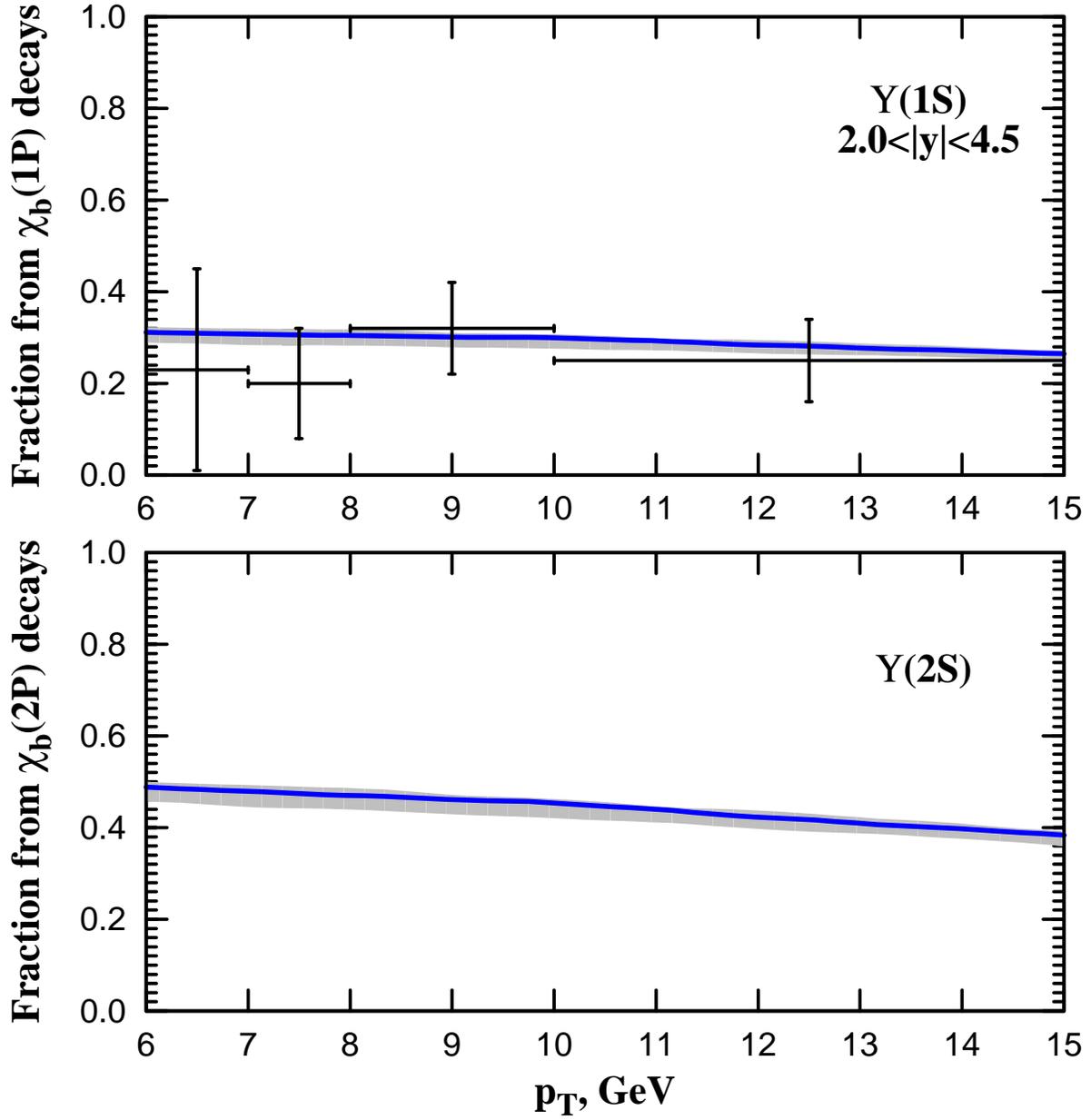}
\end{center}
\caption{ Transverse momentum distributions of
$\Upsilon(1S)$ fraction producing via $\chi_b(1P)$ decays in $pp$
scattering with $\sqrt{S}=7$ TeV and $2.0<|y|<4.5$. The data are
from the LHCb Collaboration \cite{lhcb}. }\label{figfrac}
\end{figure}

\begin{figure}[ph]
\begin{center}
\includegraphics[width=0.9\textwidth, clip=]{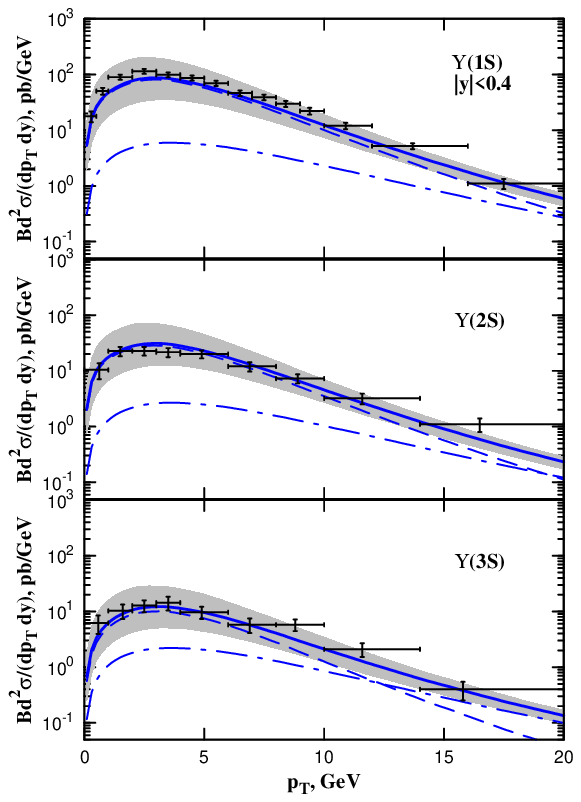}
\end{center}
\caption{ Transverse momentum distributions of prompt
$\Upsilon(1S)$, $\Upsilon(2S)$, and $\Upsilon(3S)$  hadroproduction
in $pp$ scattering with $\sqrt{S}=7$ TeV and $|y|<0.4$, including
the respective decay branching fractions
$B(\Upsilon(nS)\to\mu^+\mu^-)$. The data are from the CDF
Collaboration \cite{cdf}. The curves are the same as in the
Fig.~\ref{figatlas}.}\label{figcdf}
\end{figure}

\begin{figure}[ph]
\begin{center}
\includegraphics[width=0.8\textwidth, clip=]{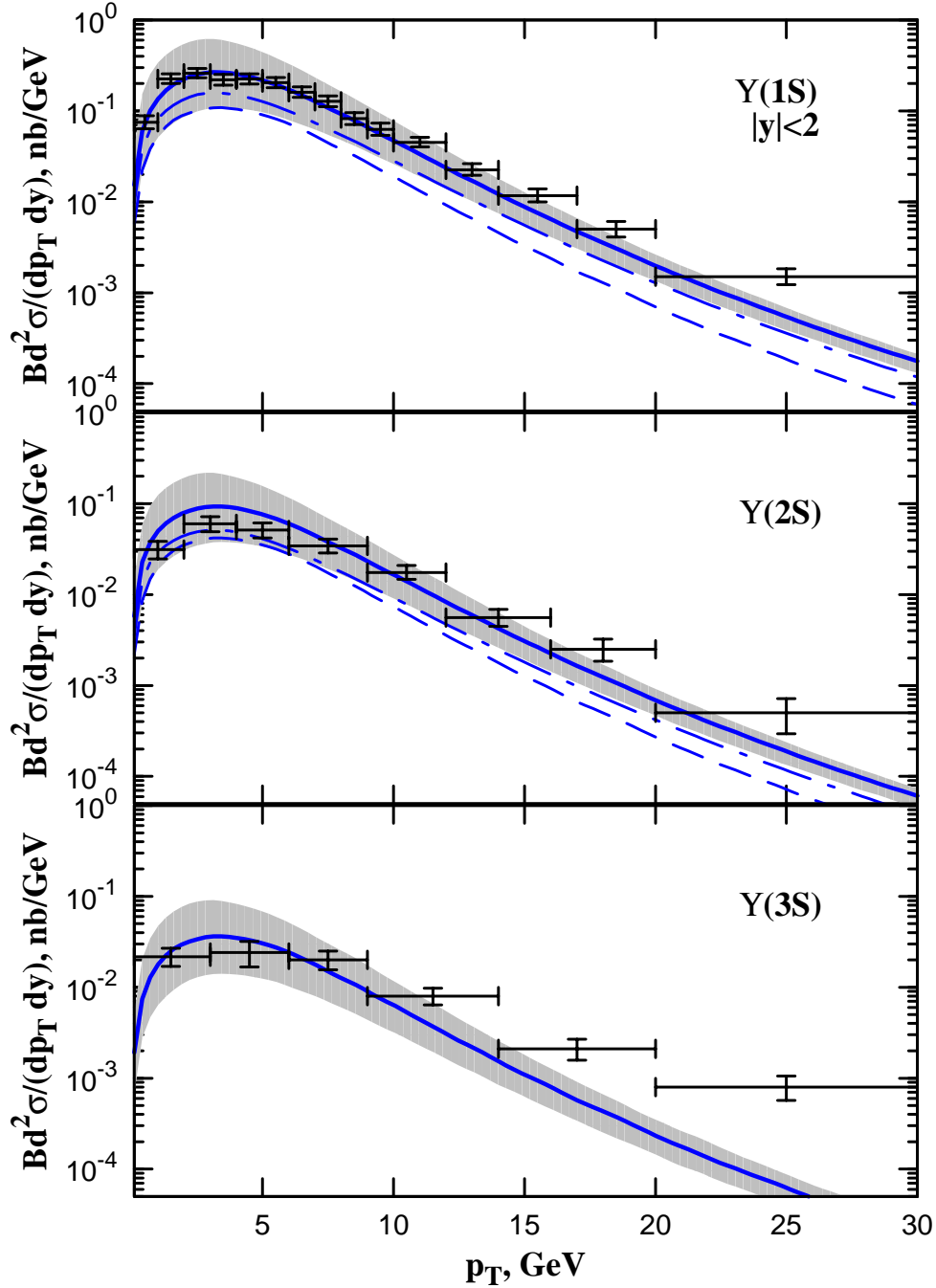}
\end{center}
\caption{ Transverse momentum distributions of prompt
$\Upsilon(1S)$, $\Upsilon(2S)$, and $\Upsilon(3S)$  hadroproduction
in $pp$ scattering with $\sqrt{S}=7$ TeV and $|y|<2$, including the
respective decay branching fractions
$B(\Upsilon(nS)\to\mu^+\mu^-)$. The data are from the CMS
Collaboration \cite{cms}. The curves correspond to LO of
color-singlet model and parton Reggeization approach: dashed line is
the direct contribution, dash-dotted line is the feed-down
contribution, solid line is their sum.}\label{figsinglet}
\end{figure}

\end{document}